\documentclass[12pt,preprint]{aastex}

\usepackage{rotating,wasysym}

\newcommand{\iso}[2]{\hbox{${}^{#1}{\rm #2}$}}
\newcommand{\msun}{\ensuremath{{\mathrm{M}_{\odot}}}}

\shorttitle{Zirconium in stardust grains}
\shortauthors{Lugaro et al.}

\begin{document}


\title{The impact of updated Zr neutron-capture cross sections and new 
asymptotic giant branch models on our understanding of the $s$ process and 
the origin of stardust}


\author{Maria Lugaro}
\affil{Monash Centre for Astrophysics (MoCA), Monash University,
Clayton VIC 3800, Australia}
\email{maria.lugaro@monash.edu}

\author{Giuseppe Tagliente\altaffilmark{1}}
\affil{Istituto Nazionale di Fisica Nucleare (INFN), Bari, Italy}
\email{giuseppe.tagliente@ba.infn.it}

\author{Amanda I. Karakas}
\affil{Research School of Astronomy and Astrophysics, Australian National University, 
Canberra, ACT 2611, Australia}
\email{amanda.karakas@anu.edu.au}

\author{Paolo M. Milazzo}
\affil{Istituto Nazionale di Fisica Nucleare (INFN), Trieste, Italy}
\email{paolo.milazzo@ts.infn.it}

\author{Franz K\"appeler}
\affil{Karlsruhe Institute of Technology, Campus North, D-76021 Karlsruhe, Germany}
\email{franz.kaeppeler@kit.edu}

\author{Andrew M. Davis\altaffilmark{2,3}}
\affil{The Department of the Geophysical Sciences, The University of Chicago, Chicago,
IL 60637, USA}
\email{a-davis@uchicago.edu}

\and

\author{Michael R. Savina\altaffilmark{2}}
\affil{Materials Science Division, Argonne National Laboratory, Argonne, IL 60439, USA}
\email{msavina@anl.gov}


\altaffiltext{1}{University of Ghent, Ghent, Belgium}
\altaffiltext{2}{Chicago Center for Cosmochemistry, USA}
\altaffiltext{3}{The Enrico Fermi Institute,
The University of Chicago, Chicago, IL 60637, USA}


\begin{abstract}

We present model predictions for the Zr isotopic ratios produced by $slow$ neutron captures in 
C-rich asymptotic giant branch (AGB) stars of masses 1.25 to 4 \msun\ and metallicities 
$Z=0.01$ to 0.03, and compare them to data from single meteoritic stardust silicon carbide 
(SiC) and high-density graphite grains that condensed in the outflows of these stars. We 
compare predictions produced using the Zr neutron-capture cross section from \citet{bao00} and 
from n\_TOF experiments at CERN, and present a new evaluation for the neutron-capture cross 
section of the unstable isotope \iso{95}Zr, the branching point leading to the production of 
\iso{96}Zr. The new cross sections generally presents an improved match with the observational 
data, except for the \iso{92}Zr/\iso{94}Zr ratios, which are on average still substantially 
higher than predicted. The \iso{96}Zr/\iso{94}Zr ratios can be explained using our range 
of initial stellar masses, with the most \iso{96}Zr-depleted grains originating from 
AGB stars of masses 1.8 - 3 \msun, and the others from either lower or higher masses. The 
\iso{90,91}Zr/\iso{94}Zr variations measured in the grains are well reproduced by the range 
of stellar metallicities considered here, which is the same needed to cover the Si composition 
of the grains produced by the chemical evolution of the Galaxy. The 
\iso{92}Zr/\iso{94}Zr versus \iso{29}Si/\iso{28}Si positive correlation observed 
in the available data suggests that stellar metallicity rather than rotation plays 
the major role in covering the \iso{90,91,92}Zr/\iso{94}Zr spread.

\end{abstract}



\section{Introduction} \label{sec:intro}

Stardust grains are tiny ($\sim \mu$m) specks of dust extracted from 
primitive meteorites, of which laboratory analysis has 
revealed isotopic compositions of many elements completely 
different from those of the bulk of solar system material. 
Large isotopic anomalies in an extended list of elements cannot be wrought by
chemical fractionation only but must be produced by nuclear reactions, which means 
that stardust 
grains carry the signature of their formation environments around 
different types of astrophysical objects, from giant stars to novae and 
supernovae. Different types of stardust grains have been recovered to 
date including diamond, graphite, silicon carbide, silicon nitride, and 
various types of oxides and silicates. 
See \citet{clayton04}, \citet{zinner08}, and \citet{davis11} for reviews,
and \citet{lugaro05} for a textbook on the topic. 

Stardust silicon carbide (SiC) grains were 
discovered in 1987 \citep{bernatowicz87} and have been the most extensively 
studied type of stardust. This is because they are easier to extract 
from meteoritic rocks than other types of grains, and 
have relatively large sizes, up to several $\mu$m, which makes them more 
manageable in the laboratory. 
Virtually all SiC grains have a stellar origin
owing to the fact that SiC can only form in C-rich gas (i.e., C$/$0$>$1),
whereas in the solar system C$/$O$\sim$0.5. 
The vast majority of stardust SiC grains 
($\sim$93\%, also referred to as ``mainstream'' SiC) originate 
from the condensation of gas into solid in 
the outer layers of the envelopes of C-rich asymptotic giant branch 
(AGB) stars of metallicity around solar and were ejected in the 
surrounding interstellar medium by strong 
stellar winds. 
Several lines of evidence point to the C-rich AGB origin for mainstream SiC 
stardust and have been extensively discussed before 
\citep[e.g.,][]{gallino90,hoppe97,lugaro99}. In summary: 
SiC molecules need a C-rich gas to form, as mentioned above, and 
their emission line at 11.2 $\mu$m
is observed in the infrared spectra of C-rich AGB stars \citep[see, e.g.,][]{speck99};
the distribution of the \iso{12}C/\iso{13}C ratio of mainstream SiC grains 
matches that of C-rich AGB stars; the Ne isotopic signature, with large 
excesses in \iso{22}Ne, is also explained as this isotope is a
main product of He burning in AGB stars. 
The distribution of Si (and Ti) shows excesses up to 20\% in the 
neutron-rich isotopes, i.e., the \iso{29,30}Si/\iso{28}Si ratios are up to 20\% higher than 
solar, and can be explained by the combined effect of the chemical 
evolution of the Galaxy, stellar migration, inhomogeneities in the interstellar medium, 
and increased condensation efficiency of 
SiC dust with increasing stellar metallicity \citep{lewis13}.
Another unmistakable signature of AGB nucleosynthesis in mainstream SiC 
grains is the presence of elements heavier than iron with isotopic 
compositions typical of the $slow$ neutron-capture process 
\citep[the $s$ process,][hereafter LDG03]{lugaro03a}. This is well known 
to occur in AGB stars as enhancements in $s$-process elements, 
such as Sr, Y, Zr, Ba, and Nd are 
spectroscopically observed \citep[e.g.,][]{smith89}. 
The 
$s$-process isotopic signature in SiC has been confirmed for several 
elements: Kr, Sr, Zr, Mo, Ba, Xe, Nd, Sm, Dy 
\citep[see, e.g.,][and 
literature therein]{nicolussi97,gallino97} and, recently, 
Eu \citep{avila13a}, W \citep{avila12a}, and Pb \citep{avila12b}. 
The 
historical discovery of the radioactive element Tc in AGB stars by 
\citet{merrill52}, which first demonstrated the occurrence of nucleosynthesis 
processes {\it in situ} in stars, 
was also confirmed by measurements of excesses in 
\iso{99}Ru in stardust SiC grains due to the radioactive decay of 
\iso{99}Tc, which is on the $s$-process path \citep{savina04}.

The origin of stardust graphite grains, on the other hand, is 
controversial and several stellar sources have been proposed. 
Low-density graphite grains show the signature of formation in the ejecta 
of core-collapse supernovae \citep[see e.g.,][]{travaglio99,pignatari13}, 
while high-density graphite grains 
appear to have originated in several different stellar sources: core-collapse 
supernovae, born-again post-AGB stars, as well as C-rich AGB stars of 
metallicity 
lower than solar \citep{jadhav08}. \citet{nicolussi98} discovered an 
$s$-process signature in the Zr composition of six high-density graphite 
grains, most likely coming from internal carbides extremely enriched 
by the $s$ process \citep{croat05}, 
strengthening the link between these grains and AGB stars.

\subsection{The $s$ process in AGB stars}\label{sec:agb}

The $s$ process occurs in the deep He-rich region of AGB stars. This region 
is usually referred to as the ``intershell'' because it is found between the 
He-burning shell, located on top of the degenerate C-O core, and 
the H-burning shell, located below the entended H-rich convective 
envelope. Hydrogen and He burning occur alternately in AGB stars. The 
H-burning shell is active most of the time, while the He-burning shell 
turns on episodically when enough H has been converted into He that 
the bottom layers of the intershell are compressed and heated up. 
Under these conditions He burning suddenly releases 
a large amount of energy, which drives convective motion in the whole intershell 
(thermal pulse, TP). This causes expansion and cooling of the whole stellar 
structure, the quenching of the H-burning shell, and, eventually, of the 
He-burning shell. At this point the star contracts again, H burning 
resumes, and a new cycle of alternate H and He burning begins 
\citep[see][for a review]{herwig05}. After each TP, a ``third dredge-up'' 
(TDU) episode may occur, which carries to the convective envelope 
(and to the stellar surface) the products of partial He burning
including \iso{12}C, \iso{22}Ne, and $s$-process elements.
For detailed models of the $s$ process in AGB stars we refer to
\citet{gallino98}, \citet{busso99}, \citet{goriely00},  
\citet{lugaro03b}, \citet{straniero06}, \citet{cristallo09}, 
\citet{bisterzo10}, and \citet{lugaro12}. Here below we report a brief description.

Two neutron sources are active in the intershell: the 
\iso{13}C($\alpha$,n)\iso{16}O and the \iso{22}Ne($\alpha$,n)\iso{25}Mg 
reactions. Neon-22 is produced via double $\alpha$-capture 
on \iso{14}N, which is very abundant in the H-burning ashes ingested in 
the TPs, and burns via ($\alpha$,n) reactions inside the convective TPs if the 
temperature reaches 300MK ($\sim$26 keV), producing a neutron flux 
over a short time (a few years) characterized by high neutron densities, 
up to 10$^{15}$ n/cm$^3$. The \iso{22}Ne neutron source is the main 
neutron source in massive AGB stars (above $\sim$4 \msun) which 
experience high temperature in their TPs 
\citep{iben77,vanraai12,karakas12,dorazi13}. In low-mass AGB stars 
(below $\sim$4 \msun) the 
\iso{22}Ne neutron source is only marginally activated and \iso{13}C 
nuclei are the neutron source. To have enough \iso{13}C nuclei to 
reproduce the observational constraints it is assumed that some mixing of 
protons occurs from the envelope into the intershell at the end of each 
TDU episode. These protons react with the abundant \iso{12}C to produce 
a region rich in \iso{13}C (the \iso{13}C {\it pocket}), 
which burns via ($\alpha$,n) reactions typically in radiative conditions 
before the onset of the next TP (i.e., during the periods in-between TPs, 
``interpulses'') at temperatures of $\sim$90MK ($\sim$8 keV), 
releasing a neutron flux over a relatively long period of time ($\sim$10$^4$ yr). 

By comparing the composition of Sr, Zr, Mo, and Ba from AGB stars and 
mainstream SiC grains, LDG03 concluded that most of these grains should 
have condensed in low-mass AGB stars. This is because the high neutron 
densities in massive AGB stars activate branching points on the 
$s$-process path and produce isotopic ratios shifted towards the 
neutron-rich isotopes, resulting in isotopic signatures that are not 
observed in the grains. This conclusion is in agreement with the fact 
that massive AGB stars experience proton-capture nucleosynthesis at the 
base of the convective envelope \citep[hot bottom 
burning,][]{boothroyd93}, which converts C into N and prevents the 
formation of C-rich gas, the necessary condition for the formation 
of SiC. For this reason, in the present paper we restrict our 
discussion to models of low-mass AGB stars, from 1.25 \msun\ to 
4 \msun, which become C-rich and where \iso{13}C is the main neutron source. 
Above this mass our models experience efficient hot bottom burning, which prevents 
them from becoming C-rich \citep{karakas10a}.

\subsubsection{Current issues with the $s$ process in low-mass AGB stars}
\label{sec:problems}

As mentioned above, many observational studies show that low-mass AGB stars are 
$s$-process enhanced and the source of free neutrons in these stars is assumed to be 
$^{13}$C nuclei. However, the mechanism for the production of the $^{13}$C nuclei is not 
well known because it depends on the treatment of convective borders in stars \citep[see 
discussion in][]{busso99}. Accurate modelling requires 3D hydrodynamical simulations of 
the interface between the envelope and the intershell of AGB stars, which are not 
available yet. In 1D AGB models a $^{13}$C pocket is artificially introduced by assuming 
the existence of some partial mixing of protons from the envelope into a thin layer at the 
top of the intershell at the end of each TDU episode. This is under the assumption 
that the mixing 
leading to the formation of the \iso{13}C pocket can occur only once the TDU has produced 
a sharp discontinuity between the convective envelope and the radiative intershell.
Free parameters allow us to adjust the features of the mixing zone in 
order to match the observations \citep[e.g.,][]{goriely00,cristallo09,bisterzo10,lugaro12}.
On top of the missing knowledge on the formation mechanism of the $^{13}$C pocket, 
there are a number of further problems associated with this $s$-process 
scenario.

\begin{enumerate}

\item{While in most cases the $^{13}$C pocket completely burns before the onset of the 
following TP, in stars of mass lower than $\sim$2 \msun, the temperature may be low enough 
that a significant fraction of $^{13}$C is left in the pocket at the time of the ingestion 
in the following TP \citep{cristallo09,lugaro12}. This $^{13}$C burns convectively inside 
the TP with the results of (i) decreasing the overall neutron exposure (i.e., the total 
number of free neutrons) due to presence of \iso{14}N ingested in the TP from the 
H-burning ashes, a strong neutron poison via the \iso{14}N(n,p)\iso{14}C reaction, and 
(ii) increasing the neutron density, due to the shorter burning timescale \citep[see 
discussion in][]{guo12}. This preferentially occurs in the first stages of the 
thermally-pulsing AGB phase, as the star heats up as it evolves. It should be noted that 
the occurrence of this $^{13}$C ingestion in the TP is strictly connected to the 
uncertainties related to the occurrence of the TDU at these very low masses: in the current 
scenario AGB stars will form a $^{13}$C pocket only if they experience the TDU and the 
lowest mass at which AGB stars experience the TDU is unknown, extremey model dependent 
\citep{frost96,mowlavi99a,stancliffe07}, and poorly constrained by observations 
\citep{wallerstein98}.}

\item{All stars rotate but the effect of rotation on the $s$ process is still very 
poorly determined. The angular velocity profile inside AGB stars may present a steep 
discontinuity between the contracting core and the expanding envelope and result in 
mixing inside the \iso{13}C pocket, which carries the \iso{14}N neutron poison into the 
\iso{13}C-rich layers and lowers the neutron exposure 
\citep{herwig03,siess04,piersanti13}. Effects such as magnetic fields 
\citep{suijs08} and gravity waves can modify the evolution of the angular 
momentum in the star and may reduce the difference in the angular velocity between the core 
and the envelope, but have not been considered in $s$-process studies so far.} 

\item{Overshoot at the bottom of the TP can lead to increased 
temperatures and the activation of the $^{22}$Ne neutron source in the low-mass AGB models, 
together with an increase of the amount of \iso{12}C in the intershell 
resulting in higher neutron exposures in the \iso{13}C pocket 
than models without overshoot \citep{herwig00,lugaro03b}.
While it has been shown that this overshoot in a 3 \msun\ star produces $s$-process 
predictions that do not agree well with observations, including constraints 
from Zr in stardust \citep{lugaro03b}, a comprehensive study is not yet available.
Recent models by \citet{pignatari13a} include such overshoot into a 1.5 \msun\ 
star, which will allow further analysis of this effect.}

\item{Finally, there is another possibility for the formation of the \iso{13}C neutron 
source where protons are ingested directly at the top of the TPs, in which case the 
neutrons for the $s$ process are all released in the convective region 
\citep{cristallo09a,lugaro12}. This process and the mass and metallicity range where it 
occurs are very uncertain as again depend on the treatment of convective borders 
in stars. In this case, first multidimensional simulations are available to guide the 1D 
models \citep{stancliffe11,herwig11,herwig13}.}

\end{enumerate}

While some progress has been achieved in the modelling of these potential effects, 
their significance on our overall understanding of the $s$ process has not yet been 
pinned down and their impact on the interpretation of observational constraints is 
unclear. In the following we will use Zr in stardust to address points 1 and 2 above,
and discuss possible implications of point 3. 
Point 4 requires parametric models of proton-ingestion episodes, which we plan as 
future work.

\subsubsection{$s$-process Zr from AGB stars and stardust grains}\label{sec:Zr}

Zr is a typical $s$-process element belonging to the first $s$-process peak in the solar 
abundance distribution \citep[predicted $s$-process contributions to its
solar abundance range from $\sim$70\% to $\sim$90\%,][]{arlandini99,goriely99,travaglio04,sneden08}.
Comparison of predictions of the isotopic abundances of Zr to stardust data are 
critical to constrain the 
$s$ process in AGB stars and to identify the range of masses and metallicities of the 
SiC parent stars. 
The path of neutron captures along the Zr isotopes has been discussed in 
detail in LDG03 (see Figure~1 and Sec.~8 of that paper) and will not
be repeated here.   
Several of the Zr isotopes have close to the magic number of neutrons 
N=50, where \iso{90}Zr has exactly N=50, which results in the relative production of 
\iso{90,91,92,94}Zr being extremely sensitive to the overall neutron exposure. On the 
other hand, 
the abundance of the remaining stable isotope, \iso{96}Zr, is determined 
by the neutron density because its production is driven 
by the activation of the branching point at the unstable \iso{95}Zr,
with a half life of 64 days (with no dependence of the temperature). 
It should be noted that the 
\iso{90,91}Zr/\iso{94}Zr ratios may also be somewhat
affected by the branching points at the unstable 
\iso{89,90}Sr and \iso{91}Y (see LDG03),  
which are uncertain given that their neutron-capture cross sections 
have been determined only theoretically.
Finally, most of the 
abundance of the element Nb is due to the radiogenic decay of the long
lived \iso{93}Zr \citep[half-life = 1.5 Myr, and down to 0.3 Myr at 300MK,][]{takahashi87}.

In LDG03 the comparison of AGB predictions to the Zr composition of single 
SiC data presented some problems: it was not possible to 
match the few grains that exhibit an extreme deficit of \iso{96}Zr, and 
the \iso{90}Zr/\iso{94}Zr and \iso{92}Zr/\iso{94}Zr ratios in some grains
were higher than predicted. LDG03 argued that 
improved measurements of neutron-capture cross sections for the Zr 
isotopes may be a key to solving these problems. Such improved estimates 
are now available as the neutron-capture cross section of all the stable Zr isotopes
and the long-lived $^{93}$Zr have been 
recently remeasured with high precision at the Neutron Time of Flight (n\_TOF) facility at 
CERN \citep[see Sec.~\ref{sec:cross} and ][]
{tagliente08a,tagliente08b,tagliente10,tagliente11a,tagliente11b,tagliente13}. 
Furthermore, the 
Zr dataset for single SiC grains has been extended by \citet{barzyk07} since the 
analysis presented in LDG03.

The aim of this paper is to 
present updated predictions for the Zr isotopic composition 
produced by the $s$ process in AGB stars resulting from the most 
recent estimates of the neutron-capture cross sections (Sect.~\ref{sec:cross}) and 
new AGB models (Sect.~\ref{sec:models}). By comparing the revised 
predictions to the composition observed in SiC and graphite grains 
we hope to reach a better understanding of the operation of the $s$ process in low-mass AGB
stars and of the mass and metallicity of the AGB parent
stars of SiC and graphite grains (Sect.~\ref{sec:results}). Our findings and future prospects 
are summarised in Sect.~\ref{sec:discussion}.

\section{Neutron-capture cross sections of the Zr isotopes} \label{sec:cross}

The Maxwellian-averaged cross sections (MACS) used 
in the calculations presented in 
LDG03 were taken from the compilation of \citet{bao00} and were based on 
experimental data that had been mainly obtained from pioneering experiments 
performed 
in the 1970s \citep{bartolome69,boldeman75,boldeman76,musgrove77,coceva79}. 
These data are in general incomplete and in some cases present large 
discrepancies.

In May 2002 a neutron time-of-flight facility (n\_TOF) became fully 
operational at CERN with the aim of making neutron capture and fission 
measurements with high accuracy over a wide energy range \citep{rubbia98,borcea03}.  
The facility delivers neutrons produced by spallation reactions induced by 
a pulsed 20 GeV/c proton beam. The main characteristic of n\_TOF is 
the high instantaneous neutron flux in combination with a low duty cycle, 
high neutron energy resolution, and low background; this allows to collect 
neutron-capture cross section data with improved accuracy and with an 
excellent signal-to-background ratio.
Given the high quality of n\_TOF, dedicated detectors were 
developed. Deuterated benzene C$_{6}$D$_{6}$ $\gamma$-ray detectors contained in a 
cylindrical low-mass carbon fibre housing \citep{plag03} have been used for 
neutron-capture measurements. These detectors are well suited for 
accurate measurements of resonance-dominated (n,$\gamma$) cross sections, 
e.g., of light and neutron-magic isotopes. The samples were kept in 
position by a carbon fibre sample changer. The mass of the experimental 
setup was reduced as much as possible and materials such as deuterium, 
carbon, oxygen, and silicon were chosen as the main constituents of the 
scintillator and the detectors because they present very low neutron-capture 
capture cross sections. 
Moreover, the analysis of n\_TOF data benefits from the performance 
of modern data acquisition techniques with fast digitizers, which allows us 
to analyze the data offline in the most flexible way, including an 
efficient pulse shape analysis for n/$\gamma$ discrimination. Other developments 
relate to the use of the well-tested and advanced R-matrix code 
SAMMY \citep{larson06}. The recent n\_TOF data present, in general, lower capture cross 
sections than previous experiments. This can be explained in terms of the more advanced 
experimental instrumentation and software for offline 
data analysis. The main improvement is certainly related to the strong 
reduction of neutron sensitivity, i.e., the background 
induced by neutrons scattered by the sample and captured in the materials 
constituting the experimental setup. 

A full campaign of n\_TOF measurements was dedicated to the Zr stable 
isotopes and the long-lived unstable isotope $^{93}$Zr. 
The neutron sensitivity of the capture setup was particularly important in 
the Zr measurements, considering the large scattering/capture ratio that 
characterizes the Zr isotopes, and the performance of the n\_TOF 
facility, in particular the high instantaneous flux, allows for neutron-capture 
measurements on radioactive samples with high accuracy.
Table~\ref{tab:cross} reports the new Zr (n,$\gamma$) 
reaction rates as a function of temperature (in keV) together with the values
recommended by \citet{bao00} for comparison.
The calculation of the MACS was carried out by folding the capture cross
section with the thermalized stellar spectra over a wide neutron energy 
range, starting from $\sim$100 eV and extending to $\sim$500 keV
to match the thermal neutron distributions at the highest
temperatures reached during shell carbon burning in massive stars. 
In this context, the n\_TOF measurements are 
limited to a few tens of keV. 
The MACS resulting from n\_TOF data analysis are composed by a first component 
calculated directly from the extracted resonance parameters and a second component
given by the JENDL calculations \citep{shibata02} at high energy. In the n\_TOF 
measured range it is possible to extract the ratio between the MACS  
calculated using n\_TOF resonance data and those presented by theoretical evaluations 
\citep{shibata02,nakagawa05}. 
To complement n\_TOF data at higher energies two different approaches are 
possible: (1) scale the evaluation by the same factor extracted in the 
n\_TOF measured range; or (2) add the contribution at high energy as given 
by the evaluations. The contribution at high energy becomes important for 
thermal energies (kT) higher than 30 keV, since the measured range extends at 
least up to 40 keV. For the scope of this paper (where the stellar temperature 
never exceeds kT$\sim$30 keV) the only 
critical isotope is $^{93}$Zr, for which it was possible to extract n\_TOF 
resonance parameters only below 8 keV of incident neutron energy. The data used 
in the present paper have been derived using approach (1) above.
More information on the measurements and data analysis prescription can be found 
in \citet{tagliente08a,tagliente08b,tagliente10,tagliente11a,tagliente11b,tagliente13}.  

\begin{figure} 
\plotone{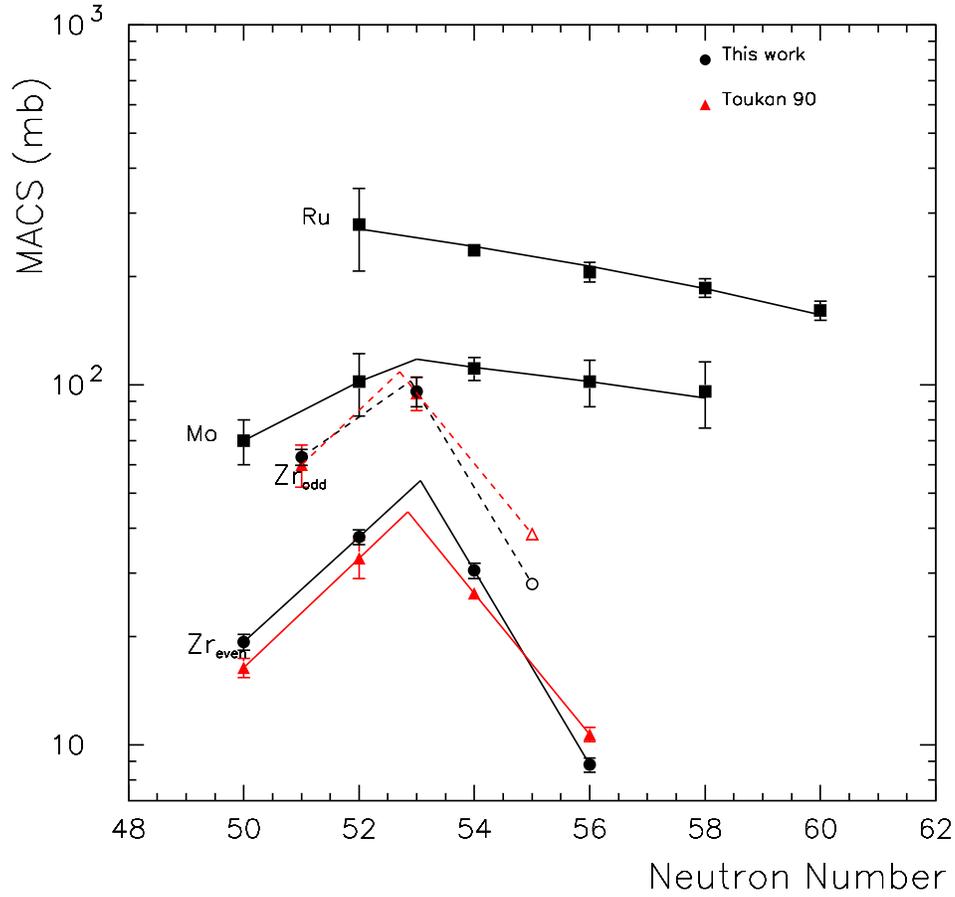}
\caption{(Color online). Systematics of the MACS data in the Zr/Mo/Ru region
used for determining the MACS of the unstable branch point isotope
$^{95}$Zr. Black points represent n\_TOF data and our current results, 
red points are from \citet{toukan90}. 
Note the strong effect of magic neutron number $N=50$.}
\label{fig:SIGMA_SYSTEMATICS}
\end{figure}

In Table~\ref{tab:cross} 
we also report our new estimate for the MACS of $^{95}$Zr. Because it is not
possible to measure the MACS of this short-lived isotope at n\_TOF, we derived it
using the method of \citet{toukan90} by applying the trend of the MACS-values
as function of neutron number for the even and odd isotopes. The method is illustrated in
Figure~\ref{fig:SIGMA_SYSTEMATICS}, which shows the strong effect of magic neutron number
N=50 on the MACS values along the isotope chains of zirconium and molybdenum.
Due to the lower MACS measured by n\_TOF for $^{96}$Zr we derived 
a $^{95}$Zr($n, \gamma$)$^{96}$Zr rate roughly 50\% lower than that reported by 
\citet{toukan90}. 
When compared to the recommended MACS of \citet{bao00}, which was obtained on the
basis of a local systematics, our result is roughly three times lower.

Finally, we note that for the neutron-captures on the Zr isotopes we are 
in the fortunate situation where the contribution of the 
laboratory-determined rate to the stellar rate is equal to unity 
\citep{rauscher12}, which means that we can safely use the laboratory 
rates as the stellar rates and apply an uncertainty equal to the 
laboratory uncertainties of $\sim$5\%, at 1$\sigma$ \citep[as compared to 
uncertainties up to $\sim$13\% reported in the][compilation]{bao00}. This is 
obviously not true for $^{95}$Zr(n,$\gamma$)$^{96}$Zr, which was instead 
derived empirically and thus carries a larger uncertainty, which is 
difficult to evaluate and could be up to a factor of two.

\section{Stellar Models} \label{sec:models}

We performed detailed nucleosynthesis calculations imposing a 
post-processing algorithm on previously computed stellar evolutionary 
sequences. The details of this procedure and the codes used to compute 
the models have been described in detail by, e.g., 
\citet{lugaro04,karakas09}.

\subsection{The stellar structure sequences}

We used stellar structure models calculated from the zero-age main 
sequence to the end of the AGB phase using the Monash Mt Stromlo stellar 
structure code \citep{lattanzio86} and included mass loss on the AGB 
phase using the prescription of \citet{vw93}. We only consider models 
that become C-rich, to allow for the formation of SiC.
We considered the 3 
\msun\ and 4 \msun\ models of metallicity $Z=0.02$ from 
\cite{karakas10a}, the 3 \msun\ model of $Z=0.01$ from 
\citet{shingles13}, and the 1.25 \msun\ and 1.8 \msun\ models of 
metallicity $Z=0.01$ from \citet{karakas10b}. The $Z=0.01$ models were 
computed with the inclusion of the C- and N-rich low temperature opacity 
tables from \citet{lederer09}, different from the 3 \msun\ and 4 \msun\ 
models of $Z=0.02$ \citep{karakas10a}. To achieve a C-rich envelope 
composition, convective overshoot was required in the 1.25 \msun\ and 
1.8 \msun\ models. We included overshoot by extending the position of 
the base of the convective envelope by $N_{\rm ov}$ pressure-scale 
heights. To obtain a C-rich envelope required $0.5 \lesssim 
N_{\rm ov} \lesssim 5$, depending on the stellar mass: in the 1.25 
\msun\ and 1.8 \msun\ models we set $N_{\rm ov} = 4$ and 3, respectively 
\citep[for more details, see][]{karakas10b}. Experimenting with the 
efficiency of overshoot to change the TDU efficiency in AGB models is 
necessary to match observational constraints such as 
the O-rich to C-rich transition luminosity in Magellanic Cloud clusters 
\citep[e.g.,][]{kamath12}, and it is justified as we still lack a reliable description 
of convective borders in stars.
This overshoot has the effect of deepening the TDU, but 
does not lead to the formation of a partially mixed zone in the top layers of the 
intershell and the consequent \iso{13}C pocket because we use instantaneous mixing in 
the evolutionary code. This is different to the time-dependent convective overshoot used 
by \citet{herwig00} and \citet{cristallo09}, which leads to the formation of the \iso{13}C 
pocket, depending on the value of the overshoot parameter $\beta$. It is only in the 
post-processing code that we artificially add a partially mixed zone into the top of 
the intershell at the deepest extent of each TDU episode to obtain the \iso{13}C pocket
(see Section~\ref{sec:nucleo}).

The main structural features of these models are presented in 
Table~\ref{tab:stars} where we report: the stellar mass (Mass, in \msun) 
and metallicity ($Z$) of the model, the number of thermal pulses followed 
by the TDU (TDUs), the number of TDUs for which C/O$>$1 is verified in the 
envelope (TDUs with C$>$O), the maximum 
temperature in the TPs (T$^{max}_{TP}$, in MK), the total mass 
dredged-up by the TDU (M$_{dred}$, in \msun), the final envelope mass 
(final M$_{env}$, in \msun), and the final C/O ratio (final C/O). We select 
AGB models that become C-rich after a certain number of pulses and do not 
suffer strong hot bottom burning. For metallicities around solar this 
corresponds to a range of masses roughly between 1.5 - 4 \msun\ 
\citep[e.g.,][]{groenewegen95,abia01,gail09}, with the lower limit 
poorly constrained, and perhaps down to 1 \msun\ \citep{wallerstein98}. 
For the 1.25 \msun, 1.8 \msun, and 3 \msun\ models of $Z = 0.01$ it was 
possible to evolve the models to very small envelope masses and to the 
end of the AGB (see Table~\ref{tab:stars}). 
In the case of the 3 \msun\ and 4 \msun\
models of $Z=0.02$, due to convergence difficulties, we 
could not evolve the star to the end of the AGB phase 
(i.e., the final envelope mass is still well above 0.01 \msun\ by the end of the 
evolution). However, since we are in a phase of very high 
mass loss ($\sim 10^{-5}$ - $10^{-4}$ \msun/yr) we do not expect any further TPs and TDUs
to occur for the 3 \msun\ model, and two further TPs and TDUs to occur for the 
4 \msun\ model \citep{karakas07b}. Table~\ref{tab:stars} also shows the 
maximum temperature achieved at the base of the TPs in each of the 
models. Recall that the \iso{22}Ne neutron source is activated only if 
this temperature reaches over 300MK. This means that this neutron source 
does not operate in the 1.25 \msun\ and 1.8 \msun\ models, it is 
only marginally active in the 3 \msun\ models, while it is at work 
in the 4 \msun\ model, though not enough to produce significant $s$-process 
enhancements without the introduction of a \iso{13}C pocket. 
   
\subsection{The stellar nucleosynthesis sequences}
\label{sec:nucleo}

The detailed $s$-process 
nucleosynthesis was calculated using a post-processing code 
that takes stellar structure information, such as temperature, density, and 
convective velocity as a function of interior mass and time, and
solves implicitly the set of equations that simultaneously describe
changes to the model abundances due to mixing and to nuclear reactions 
\citep{cannon93}. We assumed scaled-solar initial compositions, taking the solar
abundances from the compilation by \citet{asplund09}. These authors derive 
a solar metallicity of 0.0142, which we rounded to 0.015. According to
this value, our stellar structure models cover 
a metallicity range from 2/3 of solar ($Z=0.01$) to 4/3 of solar ($Z=0.02$).
We further calculated nucleosynthesis models 
with a metallicity of $Z = 0.03$ (i.e., 2 times solar) using the stellar structure
of the $Z=0.01$ (for the 1.25 \msun\ and 1.8 \msun) and $Z=0.02$ (for the 3 \msun) models, 
and changing the 
metallicity at the start of the post-processing. This resulted in a lower number of 
TDUs with C$>$O in the envelope and lower final C$/$O ratios.
This approach is not self-consistent but it is justified as a 
first approximation for small variations in the metallicity 
because the $s$-process nucleosynthesis is more sensitive to the 
metallicity (see discussion in the next section) than the evolutionary sequence is.
Comparing, for example, the stellar structure features reported in 
Table~\ref{tab:stars} for the 3 \msun\ models of $Z=0.02$ and $Z=0.01$
we see that the main difference is in the amount of TDU mass, 
which is 50\% higher in the lower metallicity model. (As mentioned above, 
the different final envelope mass is due to numerical instabilities in the 
3 \msun\ $Z=0.02$ model and we do not expect any further TDU episode for this model.)
By keeping the same stellar evolutionary sequence  
while changing the metallicity within a factor of three in the 
post-processing code we are within the model uncertainties
because the amount of TDU, which also depends on the stellar lifetime and the uncertain 
mass-loss rate, is still not well determined in AGB stellar
models \citep{frost96,mowlavi99a,stancliffe07,karakas12}.

The \iso{13}C pocket is included artificially in the 
post-processing phase by forcing the code to mix a small amount of 
protons from the envelope into the intershell at the end of each TDU.
We 
simply assume that the proton abundance in the intershell decreases 
monotonically (and exponentially) from the envelope value of $\sim$0.7 
to 10$^{-4}$ at a given point in mass located at ``M$_{mix}$'' below the 
base of the envelope. This method is described in more detail in 
\citet{lugaro04,lugaro12} and is very similar to that used by \citet{goriely00}. 
We chose M$_{mix}$=0.002 \msun,  
which produces a \iso{13}C pocket representing 
$\sim$1/10-1/20$^{\rm th}$ of the whole intershell. This allows us to reproduce the 
basic observation that the $s$-process elements in AGB stars of around 
solar metallicity are enhanced by up to an order of magnitude, with 
respect to Fe and solar abundances \citep[e.g.,][]{busso01}. 
Our results are very close to those 
obtained by \citet{cristallo09} (see also Sec.~\ref{sec:results}), 
who include the \iso{13}C pocket by introducing a 
velocity profile below the inner border of the convective envelope
and set the value of their free parameter $\beta$ to 
0.1, also in order to reproduce basic observational constraints.
In Sec.~\ref{sec:resultsagb} we present a number of test cases where we changed
the M$_{mix}$ parameter as well as the proton profile to investigate their impact
on the Zr isotopic ratios.

We employed a network of 320 nuclear species from neutrons and protons up 
to bismuth. Nuclear reaction rates were included 
using the $reaclib$ file provided by the Joint 
Institute for Nuclear Astrophysics \cite[JINA,][]{cyburt10}, as of May 
2012 (reaclib\_V2.0). The rates of the neutron source reactions correspond to 
\citet{heil08} for the \iso{13}C($\alpha$,n)\iso{16}O and to 
\citet{iliadis10} for the \iso{22}Ne($\alpha$,n)\iso{25}Mg and 
\iso{22}Ne($\alpha$,$\gamma$)\iso{26}Mg reactions. For the  
neutron-capture cross sections, the JINA reaclib database includes 
the KADoNiS database \citep{dillmann06}\footnote{We used 
the rates labelled as $ka02$ in the JINA
database (instead of $kd02$) as they provide the best fits to KADoNiS 
at the temperature of interest for AGB 
stars.}. For the Zr neutron-capture
cross sections we run models using the values from \citet{bao00} 
and from this work (Table~\ref{tab:cross}).

\section{Results and discussion} \label{sec:results}

\begin{figure} 
\epsscale{.80} 
\plotone{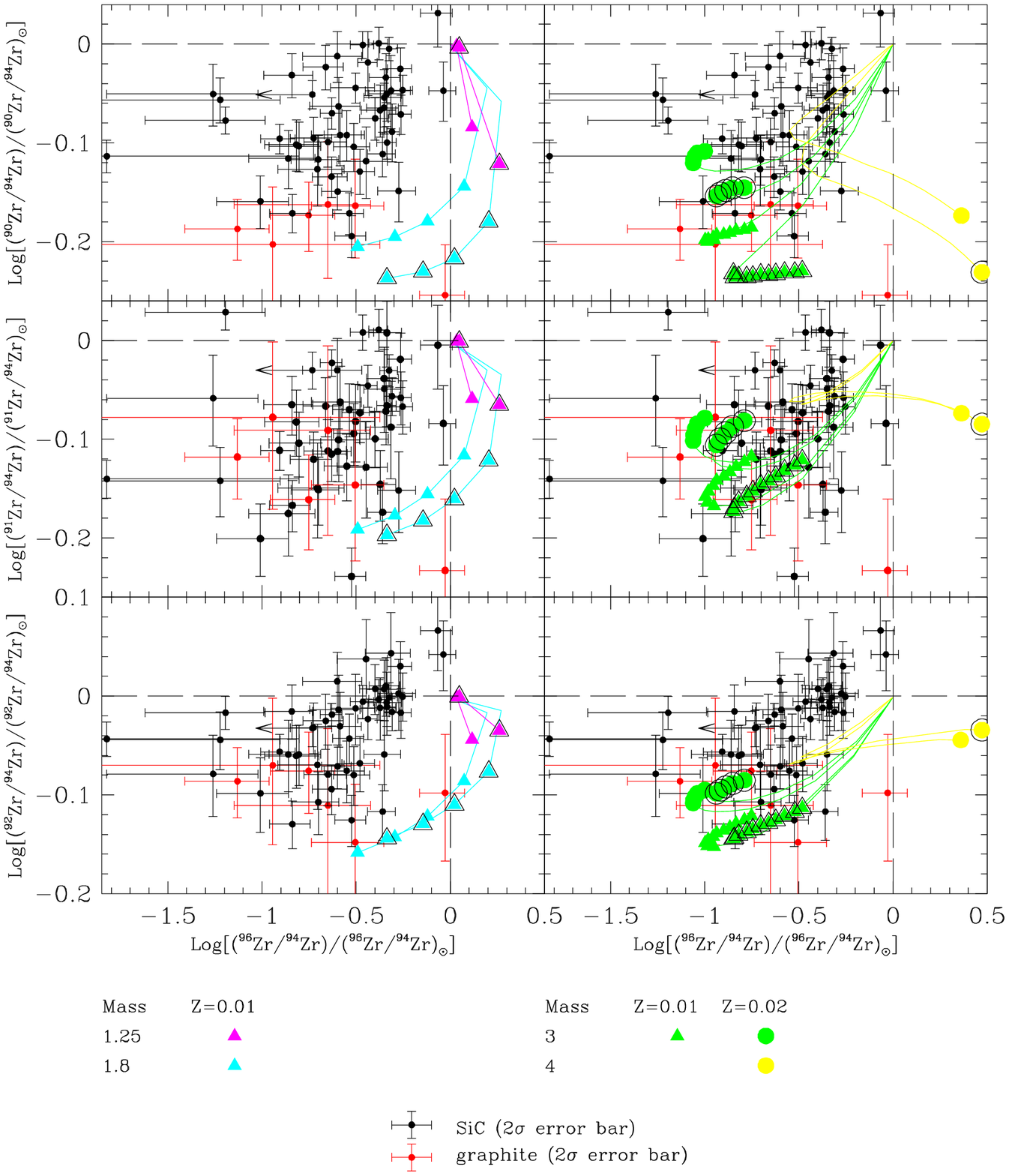} 
\caption{(Color online). Zr isotopic ratios 
measured in single SiC grains \citep{nicolussi97,davis98,davis99,barzyk07}
and high-density graphite grains \citep{nicolussi98}
and predicted by our 
AGB models (all computed with M$_{mix}$=0.002 \msun) using MACS values for the 
Zr isotopes from the present work (symbols without the black contour)
and from \citet{bao00} (symbols with the black contour). 
The AGB symbols represent the composition at the stellar surface 
after each TDU episode for which C/O$>$1.}
\label{fig:Baonew}
\end{figure}

Figures~\ref{fig:Baonew}, \ref{fig:new}, and \ref{fig:new9012} compare the Zr isotopic 
composition at the stellar surface of AGB models to the grain data. As mentioned 
above, the \iso{96}Zr/\iso{94}Zr ratio depends on the neutron density, which is mostly 
determined by the stellar mass, while the other ratios depend on the neutron exposure, which 
in our models is mostly determined by the stellar metallicity. First, we show and discuss the 
changes in model predictions driven by the updated Zr neutron-capture cross sections and 
associated uncertainties. Second, we discuss in detail the mass and metallicity dependencies 
and their implications, including the impact of the treatment of the formation of the $^{13}$C 
pocket and of stellar rotation.
 
\subsection{The impact of the new MACS}
\label{sec:macs}

Figure~\ref{fig:Baonew} presents the Zr isotopic composition at the stellar surface of 
our five AGB models computed with different sets of MACS for the Zr isotopes as 
compared to the grain data. 
The results for the 3 \msun\ models computed with the MACS from \citet{bao00} present 
similar trends to the 3 \msun\ models shown in Figure~5 of LDG03, even though 
\iso{96}Zr/\iso{94}Zr does not reach values as high as the 3 \msun\ models presented in 
LDG03 because there are less TPs with C/O$>$1 in the envelope. The main reason for this 
is the different choice of the mass-loss rate, where the mass-loss rate from 
\citet{vw93} used here typically result in less TPs than the mass-loss from 
\citet{reimers75} used in LDG03. \citep[See][for a comparison of different 3 \msun\ 
$Z=0.02$ models.]{lugaro03b,stancliffe07}.
Using the neutron-capture cross sections from \citet{bao00} 
we found similar problems as already discussed in LDG03 and mentioned in 
Sec.~\ref{sec:intro}: the handful of grains with \iso{96}Zr/\iso{94}Zr ten times lower 
than solar are not reached by the models and some of the highest 
\iso{90,91,92}Zr/\iso{94}Zr ratios are also unmatched. We cannot invoke stardust 
experimental uncertainties as the reason why the models do not cover some of the grains 
because the plotted measurement error bars are at 2$\sigma$. Updating the MACS of the 
Zr isotopes to the new values presented in Sec.~\ref{sec:cross} partly solved the 
problems above. The lower MACS for \iso{95}Zr allows our 3 \msun\ models to reach 
\iso{96}Zr/\iso{94}Zr lower than solar/10. On the other hand, even using the new 
MACS for \iso{95}Zr the 4 \msun\ model produces final \iso{96}Zr/\iso{94}Zr higher than 
solar, confirming the result of LDG03 that the grains should come from stars of mass up 
to $\sim$3 \msun. This model will be further discussed in the next section.

The lower MACS for \iso{90}Zr allows our models to reach higher \iso{90}Zr/\iso{94}Zr. For 
\iso{91}Zr/\iso{94}Zr no major change results, while for \iso{92}Zr/\iso{94}Zr the match with 
the data is slightly worse. It is still not possible to match the large fraction of grains 
with \iso{92}Zr/\iso{94}Zr around solar, unless we consider the 2$\sigma$ $\sim$10\% 
uncertainties associated with the new MACS. When multiplying the MACS by a factor between 0.9 
or 1.1 the isotopic ratios vary linearly with the change in the MACS, for example, when we 
multiply the MACS of \iso{92}Zr by 0.9 we obtain \iso{92}Zr/\iso{94}Zr 1.1 times higher (+0.04 
in the figure Log scale), when we multiply the MACS of \iso{92}Zr by 0.9 and at the same time 
multiply the MACS of \iso{94}Zr by 1.1 we obtained \iso{92}Zr/\iso{94}Zr 1.2 times higher 
(+0.08 in the figure Log scale), and so on. This rule holds for all Zr isotopes except 
\iso{96}Zr, which is much more sensitive to the MACS of \iso{95}Zr and did not show any 
significant variations when varying its MACS within 10\%.

The MACS of \iso{93}Zr is particularly interesting because it 
determines the production of monoisotopic Nb, 
given that the stable \iso{93}Nb is made by the radioactive decay of \iso{93}Zr. The value  
presented here is not significantly changed from 
\citet{bao00}, and we confirm the results of LDG03 for the elemental Zr and Nb abundances 
used by \citet{kashiv10} for comparison to the SiC data.

Finally, when considering the effect of nuclear physics inputs, it should be kept in 
mind that many other reactions that involve light elements may also have some effect on 
the neutron exposure from the \iso{13}C neutron source. This is due to changes 
in the abundances 
of the \iso{13}C nuclei and the light neutron poison nuclei (e.g., \iso{14}N),
and to the recycling of the protons produced by the 
\iso{14}N(n,p)\iso{14}C reaction, as discussed in detail by \cite{lugaro03b}. For 
example, the \iso{12}C(p,$\gamma$)\iso{13}N reaction rate  
in the JINA reaclib\_V2.0 database
is updated to \citet{li10}. This rate is up to 20\% higher than 
the NACRE rate \citep{angulo99} and produces up to 10\% lower   
\iso{90,91,92}Zr/\iso{94}Zr ratios (-0.04 in the figure Log scale), depending 
on the model, due to a higher abundance of \iso{13}C in the pocket.

\subsection{The impact of new AGB models}\label{sec:resultsagb}

\begin{figure} 
\epsscale{.80} 
\plotone{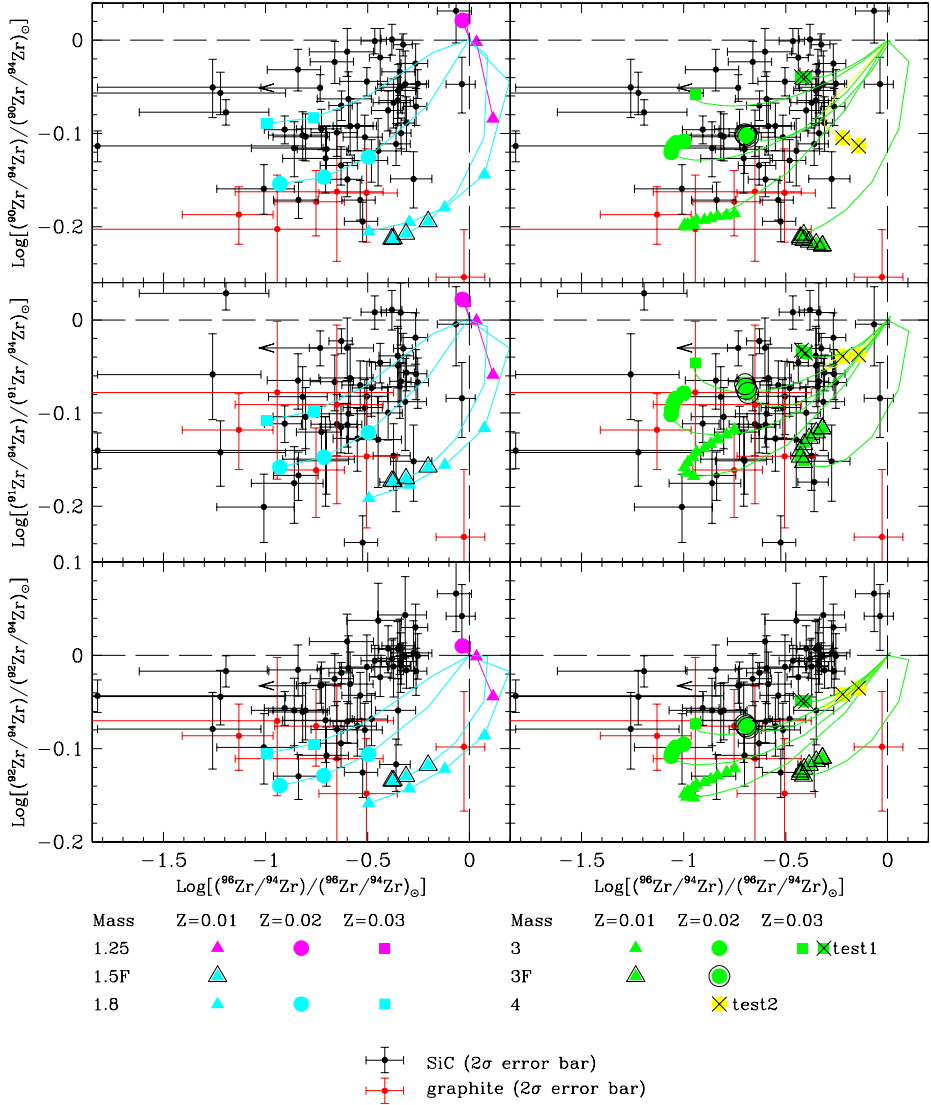} 
\caption{(Color online). Same as Figure~\ref{fig:Baonew}, except that all the models 
are computed using the Zr MACS values from this work and more AGB model predictions are 
plotted, extending the predictions to higher metallicities, and including results 
from three models from the 
FRUITY database (label ``F''). All our models were computed using M$_{mix}$=0.002 \msun, except
for the cases labelled as ``test'', which where computed with M$_{mix}$=0.0002 \msun.}
\label{fig:new}
\end{figure}

\begin{figure} 
\epsscale{.80} 
\plotone{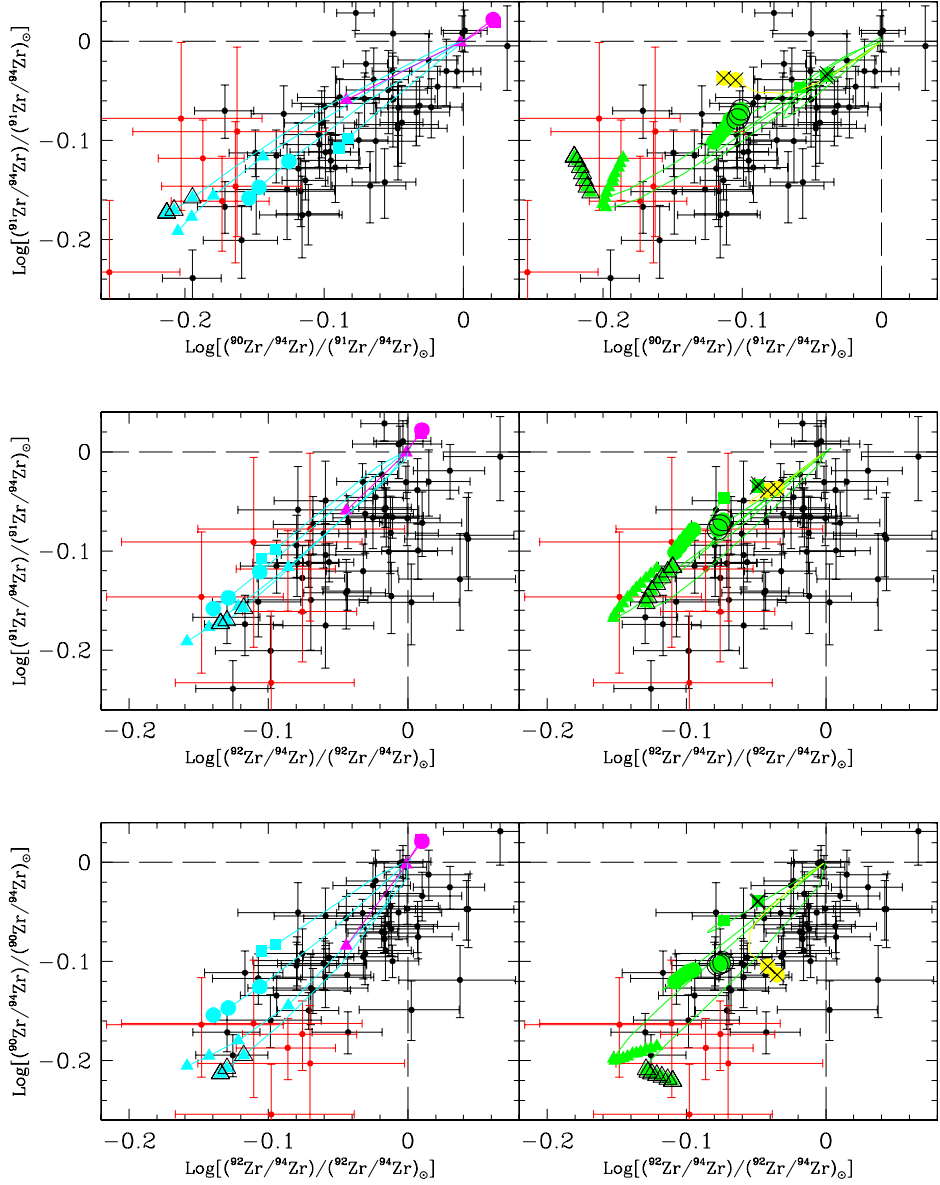} 
\caption{(Color online). Same as Figure~\ref{fig:new}, except that the isotopic ratios 
$^{90,91,92}$Zr/$^{94}$Zr are plotted against each other.}
\label{fig:new9012}
\end{figure}

Figure~\ref{fig:new} presents a number of AGB models computed with the new MACS  
in comparison to the grain data and similar models 
from the FRUITY database \citep{cristallo11}. 
Overall, we find a very good agreement between our models and the FRUITY models, with 
the differences most likely due to different MACS and to the fact that the extent 
of the \iso{13}C pocket is kept constant in our models, while it decreases with the pulse number 
in the FRUITY models \citep{cristallo09}. 
For a self-consistency check, in Figure~\ref{fig:new9012} we present the same 
data and models as in Figure~\ref{fig:new}, but plot 
$^{90,91,92}$Zr$/^{94}$Zr against each other in the three possible 
combinations. This plot highlights the composition of the unusual 
grain with $^{91}$Zr$/^{94}$Zr higher than solar 
that cannot be reached by any of the models. 

\subsubsection{The effect of the stellar mass}

As mentioned in Sec.~\ref{sec:Zr}, the \iso{96}Zr/\iso{94}Zr ratio depends on the 
activation of the \iso{95}Zr branching point. This is a function of the neutron 
density, which in turn is a function of the mass. In the 3 \msun\ and 4 \msun\ models 
\iso{96}Zr/\iso{94}Zr decreases during the first TDU episodes and then increases during 
the final TDUs due to the higher temperatures, leading to marginal 
activation of the \iso{22}Ne neutron source. In the 3 \msun\ models the last computed TP 
reaches 302MK and 305MK, for metallicities $Z=0.02$ and $Z=0.01$, respectively, and in the 4 
\msun\ model it reaches 332MK (see Table~\ref{tab:stars}), with the last 8 TPs 
experiencing temperatures in excess of 300MK in this model. The surface 
$^{24}$Mg/$^{25}$Mg ratio, which can be taken as a quantitative indicator of the 
activation of the \iso{22}Ne neutron source, in the 4 \msun\ model changes from the 
initial solar value of 7.9 to a value in the range 5.9 - 6.8, depending on the adopted 
M$_{mix}$ because \iso{14}N in the pocket adds to the \iso{22}Ne amount in the intershell, 
while in the 3 \msun\ models it changes to 6.6 - 7.2 (at $Z=0.01$) and 7.3 - 7.6 (at 
$Z=0.02$).

In the 1.25 \msun\ and 1.8 \msun\ models the opposite happens because the \iso{22}Ne 
neutron source is never activated (the $^{24}$Mg/$^{25}$Mg ratio remains solar), 
however, during the first TPs the \iso{13}C pocket is engulfed in the TPs instead of 
burning during the interpulse periods (Sec.~\ref{sec:problems}, point 1). For 
example, during the interpulse period following the first TDU episode of the 1.8 \msun\ 
model the temperature at the location of the \iso{13}C pocket reaches only 76MK. This 
behaviour has also been reported by \citet{cristallo09} for their 2 \msun\ stellar model 
at solar metallicity. The result is a higher neutron density produced by the \iso{13}C 
neutron source in the first phases of the evolution, e.g., 1.2 $\times$ 10$^9$ n/cm$^3$ in 
the TP following the first TDU episode of the 1.8 \msun\ model, which results in 
\iso{96}Zr/\iso{94}Zr higher than solar. After the first few TPs of the 1.8 \msun\ model, 
\iso{13}C burns radiatively producing low neutron densities and the \iso{96}Zr/\iso{94}Zr 
moves toward values lower than solar. In the lowest mass model presented here (1.25 \msun) 
most of the $^{13}$C nuclei burn after they are ingested in the TP, which results in 
low neutron exposures and a close-to-solar $s$-process composition.

We can conclude that the grains showing the lowest values of the $^{96}$Zr$/^{94}$Zr 
ratios (lower than solar/10) are best explained by models of masses between 1.8 \msun\ and 
3 \msun. We stress that this conclusion is possible only thanks to our updated MACS for 
\iso{95}Zr, and that unfortunately the uncertainties in the MACS for \iso{95}Zr are still 
significant. We should also keep in mind that the possible inclusion of overshoot at the 
base of the TP convective region (Sec.~\ref{sec:problems}, point 3) would lead to an 
increase of the efficiency of the \iso{95}Zr branching point \citep{lugaro03b}, shifting the 
mass range determined here. The significant number of more mildly $^{96}$Zr-depleted 
grains, with $^{96}$Zr$/^{94}$Zr $\sim$1/3 - 1/2 of solar ($-$0.5 - $-$0.3 in the figure 
Log scale), can be interpreted as a result of either (i) the activation of the \iso{22}Ne 
neutron source in stars of mass $>$ 3 \msun, or (ii) the lower $s$-process production due to 
the lower neutron exposure associated with the \iso{13}C pocket ingested in the TP in stars of 
mass between 1.25 \msun\ and 1.8 \msun. The clear lack of grains with 
$^{96}$Zr/$^{94}$Zr between solar and 1/2 of solar can potentially be used to infer the 
maximum or the minimum mass of a C-rich star within scenarios (i) or (ii), respectively. 
Case (ii) would have implications on point 1. of Sec.~\ref{sec:problems}:
because the 1.25 \msun\ model sits close to solar composition, it 
appears to be ruled out as the site of origin for the majority of the grains and we would 
need a minimum mass for C-rich stars (of $\sim$ solar metallicity) between 1.25 \msun\ and 1.8 
\msun\ in order to overcome the gap and match the grains at $^{96}$Zr/$^{94}$Zr $\sim$ 
1/2 of solar. On the other hand, the 1.25 \msun\ model represents 
a potential explanation for the two grains with $^{96}$Zr$/^{94}$Zr close to solar. A 
similar conclusion was reached by \citet{avila13b} in relation to the Ba composition of 
the extremely large ($\sim$5-20 $\mu$m) LU+LS SiC grains. In this case, compositions close 
to solar may be related to specific conditions for the formation of the lowest 
mass C-rich stars.

\subsubsection{The $^{13}$C-pocket uncertainties}

When we decreased M$_{mix}$ by a factor of 10 (i.e., M$_{mix}$=0.0002 \msun) in the 3 
\msun\ $Z$=0.03 model we found another possible solution for the grains with 
$^{96}$Zr$/^{94}$Zr) $\sim$1/3 - 1/2 of solar (``test1'' in Figure~\ref{fig:new}). Such a 
solution related to varying M$_{mix}$ would favour a stochastic process for the formation 
of the \iso{13}C pocket. However, this process would have to be fine-tuned to avoid 
producing grains with $^{96}$Zr/$^{94}$Zr between solar and 1/2 of solar, where the data 
show the clear gap discussed above.
We also note that applying the same M$_{mix}$ choice to the 4 \msun\ model we obtained 
$^{96}$Zr$/^{94}$Zr lower than solar (``test2'' in Figure~\ref{fig:new}). The difference 
between this case and the M$_{mix}$=0.002 \msun\ case, which produced $^{96}$Zr$/^{94}$Zr 
higher than solar, is due to lower amounts of \iso{14}N in the mixing zone, which is ingested 
in the TPs and converted into \iso{22}Ne. Note that if no \iso{13}C pocket is 
introduced in the 4 \msun\ model, the Zr isotopic ratios remain solar within 1\%.

We also considered models where we
investigated the effect of introducing different profiles of protons to produce the 
\iso{13}C pocket. As explained in Sec.~\ref{sec:models} in all the models 
presented so far we have assumed that the proton abundance included below the base of 
the convective envelope at the end of each TDU decreases exponentially (i.e., as 
$10^{-x}$, where $x$ is the depth in mass) from the envelope value of $\sim$0.7 to 
10$^{-4}$ at M$_{mix}$=0.002 \msun\ below the base of the envelope. 
When we changed the proton profile to follow the 
exponential of $x^{1/3}$, $x^{1/2}$, $x^2$, and $x^3$ instead of $x$ we did not find 
any significant changes in the Zr isotopic ratios for the 3 \msun\ $Z=0.02$ model.
All the variations were well within the nuclear uncertainties discussed in the 
previous section.
We then assumed that the proton 
abundance decreases exponentially starting from values ranging from 0.5 to 0.001, i.e., 
lower than the envelope value of 0.7. We found significant differences only when the 
starting value was decreased to 0.001, in which case 
$^{90,96}$Zr$/^{94}$Zr increased by 20\%, (+0.08 in the figure Log scale).

\subsubsection{The effect of stellar metallicity and rotation}

As mentioned in Sec.~\ref{sec:Zr}, the isotopic ratios involving \iso{90,91,92}Zr, 
being close to magic number 
of neutrons N=50, depend on the neutron exposure produced by 
the \iso{13}C neutron source. Because this neutron source is primary, i.e., it 
is produced starting from the H and He initially present in the star, it is 
well know that its neutron exposure $\sim$ \iso{13}C/$Z$ \citep{clayton88,gallino98}. 
Thus, varying either the amount of \iso{13}C, as done in LDG03, or 
the stellar metallicity, as done here, results in variations in the neutron 
exposure and in \iso{90,91,92}Zr/\iso{94}Zr. 
Specifically, the \iso{90,91,92}Zr/\iso{94}Zr ratios increase
by increasing the metallicity or decreasing the amount of \iso{13}C. 
We are encouraged in our approach of varying the metallicity instead of the 
amount of \iso{13}C
by the fact that the six graphite grains showing the Zr $s$-process signature are better 
matched by models of AGB stars of metallicity lower than solar, which have 
been already identified as the stellar sources of some low-density graphite grains on 
the basis of their Ne \citep{heck09} and Kr \citep{amari03} compositions. 
We also note that the 
metallicity range of the grain parent stars found here from their Zr composition 
is in agreement with that derived from their Si composition 
\citep{lewis13}. 
Furthermore, the metallicity
range considered here is well determined from the models:
stars with metallicity higher than $Z\sim$0.03 do not become C rich; while stars 
with metallicity 
lower than $Z\sim$0.01 produce the same Zr isotopic ratios as the $Z=0.01$ models
as below such metallicity the first bottleneck at the magic number of neutron of 50 is always 
bypassed and the Zr isotopic ratios reach an asymptotic behaviour.

We tested some 3 \msun\ $Z=0.02$ stellar models where we considered in a very rudimentary way 
the possibility that stellar rotation plays a major role in shaping the neutron exposure in 
the \iso{13}C pocket. As mentioned in Sec.~\ref{sec:problems} (point 2),
the potential effect of stellar 
rotation is to lower the neutron exposure \citep{herwig03,siess04,piersanti13}. Clearly, 
changing the initial velocity of any given stellar model, as well as considering the 
effect of, e.g., magnetic fields, would allow for a wide range of possibilities. We simulated 
them simply by allowing a flat proton profile mixed in the intershell region down to 
M$_{mix}$=0.002 \msun\ below the base of the convective envelope and tested a wide range of 
values for the constant assumed number of protons, from 10$^{-4}$ to 0.05. These tests allowed 
us to cover a similar range of the Zr isotopic ratios predicted by our whole set of models with 
just one mass and one metallicity, in a similar way as obtained by LDG03 varying the 
\iso{13}C-pocket efficiency. 
We also note that the higher neutron exposures resulting from overshoot at the base of 
the convective pulse (Sec.~\ref{sec:problems}, point 3) would require rotation 
to reproduce the observed spread 
\citep{herwig03}. The question is how can we discriminate between the effect of 
metallicity and the effect of rotation and determine which is the primary effect in shaping 
the stardust Zr distribution?

A way forward to answer this question is provided by stardust when considering its 
\iso{29,30}Si/\iso{28}Si ratios. These increase with the stellar metallicity 
\citep[Figure 2 of][]{lewis13} because they are mostly determined by the initial 
composition of the star, which in turn is a function of metallicity via the chemical 
evolution of the Galaxy. However, they do not depend on stellar rotation. This means 
that possible positive correlations between the \iso{90,91,92}Zr/\iso{94}Zr ratios and the 
\iso{29,30}Si/\iso{28}Si ratios can provide us with a quantitative proxy for the effect 
of metallicity. Note that we do not expect strong correlations between \iso{96}Zr/\iso{94}Zr 
and any of \iso{29,30}Si/\iso{28}Si because \iso{96}Zr/\iso{94}Zr does not strongly 
depend on the metallicity (see Figure~\ref{fig:new}). The effect of rotation can then 
be derived considering the spread of \iso{90,91,92}Zr/\iso{94}Zr ratios around such 
possible positive correlations, though it should be kept in mind that these  
correlations are also 
smeared out in \iso{29,30}Si/\iso{28}Si by the effect of possible small inhomogeneities 
in the interstellar medium, which can shift the initial 
$\delta$(\iso{29,30}Si/\iso{28}Si) by $\sim70\permil$ for any given metallicity 
\citep{lugaro99,nittler05}.
 
\begin{figure} 
\epsscale{.80} 
\plotone{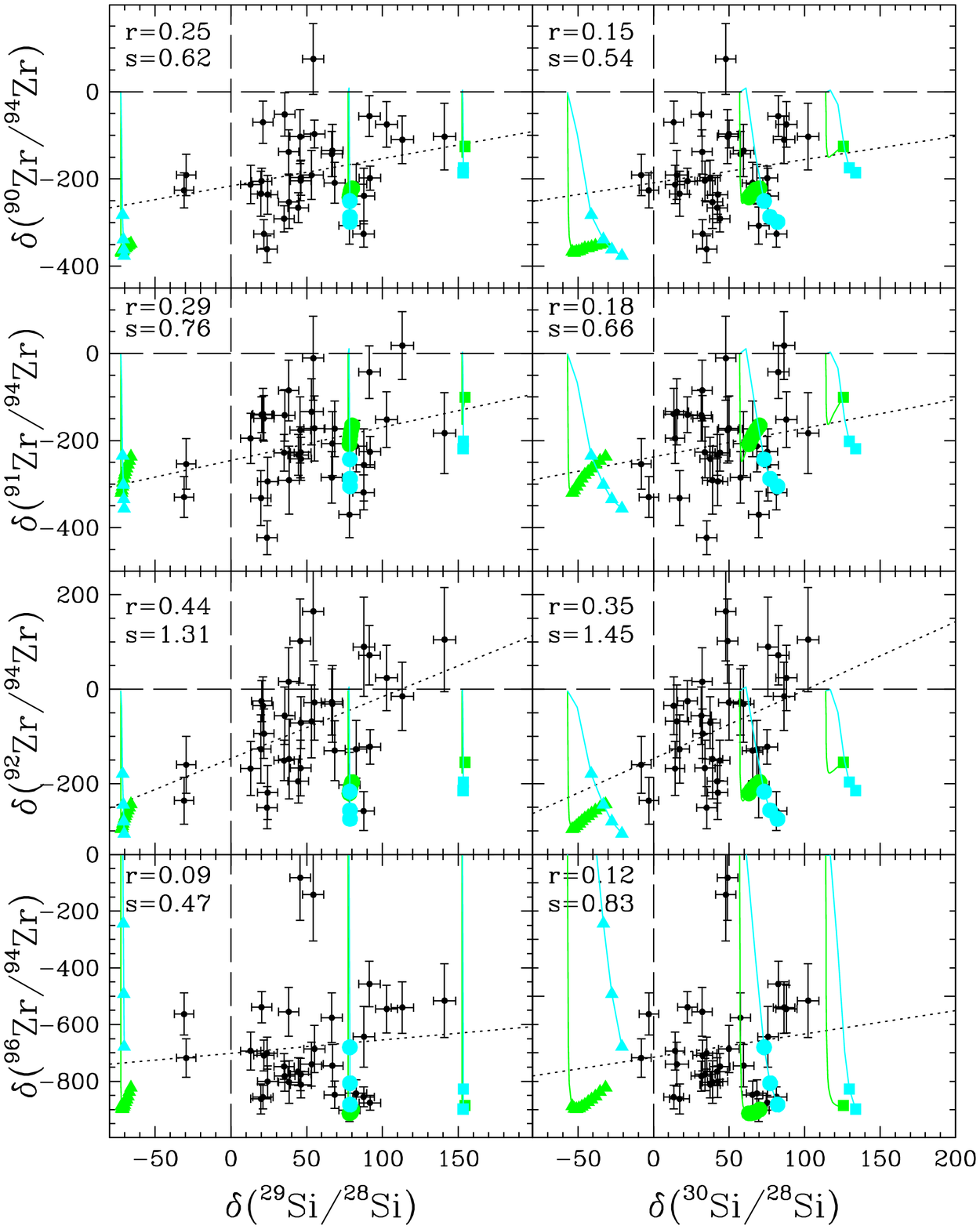} 
\caption{(Color online). All possible combinations of $\delta$(\iso{90,91,92,96}Zr/\iso{94}Zr) 
versus $\delta$(\iso{29,30}Si/\iso{28}Si) 
from the SiC dataset of \citet{barzyk07} (black symbols with 
2$\sigma$ error bars). 
The data linear regression lines 
are plotted as dotted lines, with their slopes and correlation coefficients 
indicated in each panel as ``s='' and ``r='', respectively.
The 1.8 \msun\ (cyan symbols) 
and 3 \msun\ (green symbols) models
of metallicities $Z=0.01$ (triangles), 0.02 (circles), and 0.03 (squares)
are also plotted, where we assumed 
different initial $\delta$(\iso{29,30}Si/\iso{28}Si) 
values for each different metallicity to 
account for the chemical evolution of the Galaxy \citep{lewis13}.}
\label{fig:ZrSi}
\end{figure}

In Figure~\ref{fig:ZrSi} we 
plot the Zr ratios as function of the Si ratios for the 32 available single SiC grains from 
\citet{barzyk07} and indicate their 
correlation coefficients. In this plot we use the $\delta$ notation where 
$$
\delta(^{i}{\rm Zr}/^{94}{\rm Zr}) = 
(((^{i}{\rm Zr}/^{94}{\rm Zr})/(^{i}{\rm Zr}/^{94}{\rm Zr})_\odot)-1)\times1000) 
$$
represents the permil variation of the given ratio with respect to the 
solar ratio (so that $\delta$=0 represents solar ratios and, e.g., 
$\delta=+500\permil$ means a ratio 50\% higher than solar). 
For comparison, we also plot our AGB predictions for 1.8 \msun\ and 
3 \msun\ stars and different metallicities with initial Si isotopic ratios shifted to 
account for the chemical evolution of the Galaxy as reported by \citet{lewis13}. 

The positive 
correlations produced
by the effect of metallicity are hinted at by the small set of currently available data,
particularly in \iso{92}Zr/\iso{94}Zr versus \iso{29}Si/\iso{28}Si, 
which suggests that the effect of metallicity is predominant.
That a correlation should be more evident in \iso{92}Zr/\iso{94}Zr versus 
\iso{29}Si/\iso{28}Si is expected because 
\iso{30}Si/\iso{28}Si is known to be more affected by AGB nuclosynthesis, as shown 
in the figure \citep[see also][]{zinner06}, 
and the 
\iso{90,91}Zr/\iso{94}Zr ratios may also be changed by the branching points at the unstable 
\iso{89,90}Sr and \iso{91}Y. It seems difficult 
to disentagle the effect of inhomogeneities
in the interstellar medium on \iso{29}Si/\iso{28}Si from the 
possible secondary effect of rotation on \iso{92}Zr/\iso{94}Zr in producing 
the spread around the regression line. Future work may involve superposing a random
choice of initial \iso{29}Si/\iso{28}Si, as expected by inhomogeneities. Also, as 
already noted in Sec.~\ref{sec:macs}, the  
\iso{92}Zr/\iso{94}Zr ratios are not well fitted by the models. This point will need to
be reinvestigated in the light of new determinations of the MACS of \iso{92}Zr.

\section{Conclusions} \label{sec:discussion}

We have compared the Zr isotopic composition derived from stellar models of 
C-rich AGB stars of 
masses 1.25 \msun\ to 4 \msun\ and metallicities 0.01 to 0.03 including updated MACS for 
the Zr isotopes to the composition of Zr measured in stardust SiC and graphite grains. Our 
main conclusions can be summarised as follows:

\begin{enumerate}

\item{The new Zr MACS measurements and the new evaluation of the MACS of the unstable branching 
point nucleus $^{95}$Zr allow a good match to the SiC and graphite data, within the nuclear 
uncertainties, except that the predicted \iso{92}Zr/\iso{94}Zr are, on average, still 
outstandingly lower than the data points. To address this problem 
new measurements of the MACS of \iso{92}Zr have been performed at 
n\_TOF and at the linear electron accelerator facility GELINA (Belgium) and the data are 
currently being analysed. Furthermore, new measurements for the MACS of \iso{93}Zr 
for a wide
energy range have been be carried out at n\_TOF and these data are also currently being analysed, 
which will allow avoiding 
the use of theoretical models for a more accurate determination of the MACS of this isotope.} 
 
\item{From analysis of the \iso{96}Zr/\iso{94}Zr ratios, we confirm the results of 
LDG03 that stellar masses below 4 \msun\ are the best candidates as the origin of the vast 
majority of the mainstream SiC grains and conclude that the most \iso{96}Zr-depleted 
grains must originate from stars of mass $\sim$1.8 - 3 \msun. However, we cannot 
unambiguously
attribute a mass to the more mildly \iso{96}Zr-depleted grains as they can be explained by 
both higher and lower masses. Measurements of other isotopic ratios affected by branching 
points (such as \iso{86}Kr/\iso{82}Kr and \iso{134}Ba/\iso{136}Ba) in the same grain will 
provide independent constraints to settle this question. Because possible overshoot at the 
base of the convective TP would change the picture outlined here, more models are needed 
considering this effect. Our lowest mass C-rich stellar model (1.25 \msun) may 
be a suitable site of origin for the two grains with close-to-solar Zr composition 
\citep[see also][]{avila13b}}.

\item{We find that the spread in neutron exposures needed to produce the range of observed 
$^{90,91,92}$Zr/$^{94}$Zr isotopic ratios can be naturally produced by considering the 
same range of metallicities ($Z$=0.01 - 0.03) needed to explain the Si isotopic 
composition of the grains, as derived from Galactic chemical evolution models 
\citep{lewis13}. If this interpretation is correct, other effects, such as stochastic 
variations in the proton profile that leads to the formation of the \iso{13}C pocket or 
stellar rotation, would play a secondary role. To verify this point we investigated 
the correlations between the $^{90,91,92}$Zr/$^{94}$Zr isotopic ratios and the 
$^{29,30}$Si/$^{28}$Si isotopic ratios in the available 32 data points from 
\citet{barzyk07} and found a hint that metallicity is the predominant effect.}

\end{enumerate}

Other elements measured in SiC may be expected to present similar correlations as the Zr and 
Si isotopic ratios discussed in the last item above. Specifically, ratios involving nuclei 
with magic numbers of neutrons, such as \iso{88}Sr/\iso{86}Sr, \iso{138}Ba/\iso{136}Ba, and 
\iso{208}Pb/\iso{204}Pb, should correlate with isotopic ratios of elements affected by the 
chemical evolution of the Galaxy, such as Si and Ti. Ba and Si ratios are available for 20 
grains from \citet{barzyk07}. These, together with 
two new Ba/Si studies of over 100 grains to be published shortly by two of us (Savina and  
Davis) will be considered in a forthcoming paper.
\citet{marhas07} analysed Ba and Si in another 16 grains, however, these analysis were 
performed 
with the NanoSIMS instrument \citep{zinner01} and may suffer from molecular interferences 
\citep{avila13a}. 
While the number of grains with both Sr, Zr, or Ba and Si or Ti data is at present quite 
limited, measurements of Sr, Ba, and Si isotopes in a suite of grains are currently underway 
using the Resonance Ionization Mass Spectrometer CHARISMA \citep{savina03b} and the 
NanoSIMS.
Furthermore, thanks to the upcoming CHicago Instrument for Laser Ionization (CHILI) 
at the 
University of Chicago \citep{stephan13} much more high-precision data will become available in 
the near future, which will be fundamental to confirm if the spread in the Zr isotopic ratios 
in SiC and graphite is/is not primarily due to a metallicity effect. At the same time, 
asteroseismology observations of white dwarfs \citep[e.g.,][]{charpinet09} and red giant stars 
\citep{mosser12} are providing us with evidence that the cores of red giant and AGB stars spin 
much more slowly than expected by simple models of the evolution of the angular momentum in 
stars and that strong coupling between the core and the envelope is required to match the 
observations \citep{suijs08,tayar13}. A slower rotating core would also lead to a smaller 
impact of rotation on the neutron exposure in the \iso{13}C pocket. More quantitative studies 
are required, which, together with the upcoming extended sample of grain data from CHILI and 
CHARISMA, 
will set firm constraints on the operation of the $s$-process in AGB stars.

\acknowledgments

We thank Mark van Raai and Robin Humble for support on the post-processing code. We thank 
Peter Hoppe for discussion on grain data. We acknowledge the constructive criticism of 
the anonymous referee, who greatly helped us to improve the structure, focus, and clarity of 
the paper. ML and AIK are grateful for the support of the NCI National Facility at the ANU. 
ML an ARC Future Fellow (supported by grant FT100100305), AIK 
is an ARC Future Fellow (supported by grant FT10100475). This work was partially supported by 
the National Aeronautics and Space Administration, through grants to AMD and MRS. The CHARISMA 
instrument at Argonne National Laboratory is supported by the US Dept. of Energy, BES − 
Division of Materials Science and Engineering, under contract DEAC02-06CH11357.






\begin{center}
\begin{table}
\caption{Reaction rates N$_{\rm A}<\sigma v>$ of (n,$\gamma$) reactions of 
the stable Zr isotopes and the unstable $^{93}$Zr and $^{95}$Zr, all given in units of 
10$^6$ cm$^3$ mole$^{-1}$ s$^{-1}$. Uncertainties are $\sim$5\% at 1$\sigma$ 
for all the isotopes, except $^{95}$Zr, whose 
cross section is derived empirically and has a larger associated uncertainty.
\label{tab:cross}}
\vspace{0.5cm}
  \begin{tabular}{ c c c c c c c c c c c c}
    \hline\hline
          \multicolumn{12}{ c }{\iso{90}Zr}\\ \hline
kT (keV) & 5 & 10 & 15 & 20 & 25 & 30 & 40 & 50 & 60 & 80 & 100 \\
Bao et al. & 2.54 & 2.84 & 2.96 & 2.95 & 3.03 & 3.04 & 3.00 & 3.17 & 3.27 & 3.54 & 3.69 \\
This work & 2.62 & 2.61 & 2.64 & 2.68 & 2.73 & 2.79 & 2.85 & 2.89 & 2.92 & 2.93 & 2.90 \\
\hline\hline
          \multicolumn{12}{ c }{\iso{91}Zr}\\ \hline
kT (keV) & 5 & 10 & 15 & 20 & 25 & 30 & 40 & 50 & 60 & 80 & 100 \\
Bao et al. & 15.0 & 12.3 & 10.7 & 9.80 & 9.12 & 8.68 & 8.18 & 8.03 & 7.97 & 8.02 & 8.44 \\
This work & 14.0 & 12.0 & 10.8 & 10.1 & 9.50 & 9.11 & 8.51 & 8.21 & 7.77 & 7.79 & 7.39 \\
    \hline\hline
          \multicolumn{12}{ c }{\iso{92}Zr}\\ \hline
kT (keV) & 5 & 10 & 15 & 20 & 25 & 30 & 40 & 50 & 60 & 80 & 100 \\
Bao et al. & 7.74 & 6.43 & 5.63 & 5.20 & 4.89 & 4.77 & 4.67 & 4.85 & 5.10 & 5.42 & 6.06 \\
This work & 8.00 & 6.72 & 6.02 & 5.68 & 5.51 & 5.46 & 5.54 & 5.71 & 5.91 & 6.35 & 6.78 \\
    \hline\hline
          \multicolumn{12}{ c }{\iso{93}Zr}\\ \hline
kT (keV) & 5 & 10 & 15 & 20 & 25 & 30 & 40 & 50 & 60 & 80 & 100 \\
Bao et al. & 20.1 & 17.8 & 16.6 & 15.6 & 14.6 & 13.8 & 12.5 & 11.6 & 11.1 & 10.4 & 9.78 \\
This work  & 20.0 & 17.9 & 16.5 & 15.5 & 14.5 & 14.1 & 13.5 & 11.9 & 12.2 & 11.6 & 11.3 \\
    \hline\hline
          \multicolumn{12}{ c }{\iso{94}Zr}\\ \hline
kT (keV) & 5 & 10 & 15 & 20 & 25 & 30 & 40 & 50 & 60 & 80 & 100 \\
Bao et al. & 4.37 & 4.34 & 4.09 & 3.90 & 3.83 & 3.76 & 3.84 & 4.28 & 4.69 & 4.95 & 5.53 \\
This work & 4.15 & 4.36 & 4.32 & 4.31 & 4.34 & 4.41 & 4.57 & 4.75 & 4.97 & 5.33 & 5.70 \\
    \hline\hline
          \multicolumn{12}{ c }{\iso{95}Zr}\\ \hline
kT (keV) & 5 & 10 & 15 & 20 & 25 & 30 & 40 & 50 & 60 & 80 & 100 \\
Bao et al. & 17.5 & 15.4 & 13.9 & 12.9 & 12.0 & 11.4 & 10.5 & 10.1 & 9.62 & 9.21 & 8.98 \\
This work  & 15.1 & 8.94 & 6.75 & 5.47 & 4.65 & 4.11 & 3.57 & 3.03 & 3.05 & 2.92 & 2.86 \\
    \hline\hline
          \multicolumn{12}{ c }{\iso{96}Zr}\\ \hline
kT (keV) & 5 & 10 & 15 & 20 & 25 & 30 & 40 & 50 & 60 & 80 & 100 \\
Bao et al. & 3.31 & 2.34 & 1.95 & 1.65 & 1.58 & 1.55 & 1.39 & 1.27 & 1.19 & 1.04 & 0.93\\
This work & 3.13 & 2.18 & 1.77 & 1.52 & 1.37 & 1.28 & 1.21 & 1.21 & 1.25 & 1.34 & 1.45 \\
    \hline
    \hline
  \end{tabular}
  \end{table}
\end{center}


\begin{center}
\begin{table}[h]
\caption{Details of stellar models \label{tab:stars}}
\vspace{0.5cm}
  \begin{tabular}{ c c c c c c c c c }
    \hline\hline
    Mass & $Z$ & TDUs & TDUs & 
T$^{max}_{TPs}$ & M$_{dred}$ & final M$_{env}$ & final \\
    \msun\ & & & with C$>$O & 
MK & \msun\ & \msun\ & C/O \\
\hline
    1.25 & 0.01 & 3 & 2$^{a}$ & 246 & 0.013 & 0.026 & 2.23$^{a}$ \\
    1.8 & 0.01 & 6 & 4$^{b}$ & 266 & 0.041 & 0.014 & 3.12$^{b}$ \\
    3 & 0.01 & 16 & 11 & 306 & 0.120 & 0.004 & 3.32 \\
    3 & 0.02 & 16 & 5$^{c}$ & 302 & 0.081 & 0.676 & 1.44$^{c}$ \\
    4 & 0.02 & 15$^{d}$ & 2$^{d}$ & 332 & 0.056 & 0.958 & 1.13$^{d}$ \\
\hline
    \hline
  \end{tabular}\\

$^{a}$Using $Z=0.03$ in the post-processing we obtained
1 TDUs with C$>$O and final C/O=1.04. 

$^{b}$Using $Z=0.03$ in the post-processing we obtained
2 TDUs with C$>$O and final C/O=1.33. 

$^{c}$Using $Z=0.03$ in the post-processing we obtained
1 TDUs with C$>$O and final C/O=1.08. 

$^{d}$This model experienced mild hot bottom burning (with a temperature 
of 23MK at the base of the convective envelope), which delayed the formation of a C-rich 
envelope. As discussed in the text, this model may experience two more TDU episodes,
further increasing the C/O ratio.

  \end{table}
\end{center}

\newpage 



\end{document}